\documentclass[]{article}
\usepackage[a4paper, total={5.5in, 9in}]{geometry}
\usepackage{mathtools}
\usepackage{amsmath}
\usepackage{caption}
\usepackage{float}
\usepackage{subcaption}
\usepackage{graphicx}
\title{Testing the dynamical stability and validity of generalized second law within the phantom dynamical dark energy model}
\author{Naseeba.K.M$^1$\footnote{nasikm2001@gmail.com}, Sarath Nelleri$^2$\footnote{sarathn@iitk.ac.in} and Navaneeth Poonthottathil$^2$\footnote{navaneeth@iitk.ac.in} \\  \small $^1$Department of Physics, Government Brennen College, Dharmadam, Thalassery 670106, India \\ \small $^2$Department of Physics, Indian Institute of Technology, Kanpur 208016, India}
\date{}
\begin{document}

\maketitle

\begin{abstract}
Hubble constant($H_0$) tension and tension in the matter fluctuation amplitude ($s_8$) are fascinating puzzles in cosmology nowadays. Phantom dynamical dark energy model (PDDE), also known as little sibling of the big rip is an abrupt event that can happen in the far future evolution of the universe. Recent analysis of PDDE model based on CMBR data shows that the model is a potential candidate to alleviate these tension problems. In this work, we study the background evolution of the universe within the PDDE model. Analysis based on the SNIa+BAO+OHD data shows that the model is successful in explaining the late phase acceleration of the universe. Also, the values of the cosmological parameters predicted by PDDE model are consistent with the values predicted by the $\Lambda$CDM model. However, most of the phanton dark energy models doesn't give stable solution in the asymptotic future. In this regard, we address the dynamical stability of the PDDE model and also test the validity of the generalized second law (GSL) of thermodynamics. We show that the model is dynamically unstable and violates the GSL. The model doesn't satisfy the convexity condition and hence the universe doesn't behave like an ordinary macroscopic system within the PDDE model.
\end{abstract}

\section{Introduction}
Latest observational probes such as the Type Ia supernovae (SNIa), the cosmic microwave background Radiation (CMBR), the large-scale structure (LSS) in the distribution of galaxies and baryon acoustic oscillation (BAO) revealed that the present expansion of the universe is accelerating\cite{perlmutter1999measurements, riess1998observational, aghanim2020planck, tegmark2004cosmological, spergel2003first, smoot1992structure, ade2014planck, eisenstein2005detection, ade2016planck}. It requires an extraordinary energy component dubbed dark energy to explain the observed acceleration\cite{peebles2003cosmological}. The nature and origin of dark energy are still unknown. However, the cosmological constant ($\Lambda$) having equation of state parameter $\omega_{\Lambda} = -1$ could explain the late time acceleration\cite{peebles1993principles}. This model, which is in concordance with most of the observations in cosmology, is known as the $\Lambda$CDM model. However, many physicists criticize the model due to a statistically significant tension observed between the value of Hubble constant, $H_0 = 67.37\pm 0.54$ $\text{km}$ $\text{s}^{-1}$ $\text{Mpc}^{-1}$ obtained from the CMBR data assuming the $\Lambda$CDM model\cite{aghanim2020planck} and the one obtained from the Cepheid calibrated type Ia supernovae, $H_0 = 73.2\pm 1.3$ $\text{km}$ $\text{s}^{-1}$ $\text{Mpc}^{-1}$\cite{riess2021cosmic, riess2019large}. Another issue is the observed tension between the value of the parameter that quantifies the amplitude of matter fluctuation ($S_8 = 0.832\pm 0.013$) inferred from the CMBR data assuming the $\Lambda$CDM model\cite{aghanim2020planck} and one obtained from the LSS data ($S_8 = 0.762^{+0.025}_{-0.024}$)\cite{joudaki2020kids+}. Both of these disparities lead physicists to go beyond the standard model of cosmology.

Numerous theoretical models have been proposed to tackle these issues. Some of these models modify the curvature term in the Einstein field equation known as modified gravity models, see ref.\cite{odintsov2021analyzing, schiavone2023f, adil2021late, nojiri2022integral, di2021realm}. Other favourites include models that modify the energy-momentum tensor of the field equations known as modified dark energy models\cite{pan2019interacting, karwal2022chameleon, murgia2021early, niedermann2020resolving}. Most modified dark energy models are just an extension of the $\Lambda$CDM model. For instance, dynamical dark energy models that include interaction between dark energy and dark matter\cite{peracaula2021running} and early dark energy which behaves like a cosmological constant in the very early universe\cite{poulin2019early} are a potential candidate to alleviate the $H_0$ tension. Recently, phantom dark energy models, where the equation of state parameter less than -1, have gained much attention. In ref.\cite{li2019simple}, Li and Shafieloo introduced the phenomenological emergent dark energy model with zero degrees of freedom and showed that it could resolve the Hubble tension and its observational significance is as competitive as the $\Lambda$CDM model\cite{rezaei2020bayesian}.

The phantom dynamical dark energy model that smoothed the big rip singularity was presented in ref.\cite{bouhmadi2015little}. In this model, the Hubble parameter and the scale factor diverge, but the first derivative of the Hubble parameter doesn't blow up and hence the name,  little sibling of the big rip; the event happens at an infinite cosmic time. In addition, the model was studied by Bouali et al., who put constraints on the model parameters, which are consistent with the observation\cite{bouali2019cosmological}. Further extension of this model, including interaction, is presented in ref.\cite{bouali2021cosmological}. Recently, Dahmani et al.\cite{dahmani2023smoothing} studied this model extensively using the latest observational probes such as Planck18, Pantheon, and BAO data sets. The analysis shows that the model can alleviate the $H_0$ and $S_8$ tensions to $3\sigma$ and $2.6\sigma$ respectively. 

In this work, we constrain the PDDE model parameters using the data set combination SNIa+BAO+OHD and analyze the late phase background evolution of the universe. We also compare the PDDE model predictions with the $\Lambda$CDM predictions. Note that even if the phantom dark energy models are successful in predicting the late phase acceleration of the universe, a larger class of models give unstable solutions in the far future. In this regard, we perform the dynamical system analysis to test the dynamical stability of the model. We also test the validity of the generalized second law of thermodynamics and entropy maximization within the PDDE model. 

The paper is structured as follows. In the next section, we present the phantom dynamical dark energy model. In sec. 3, we constrain the model parameters using observational data and discuss the significance of each parameter and its cosmological implications. In sec. 4, we study the evolution of cosmographic parameters. In sec. 5, we perform the dynamical system analysis to test the dynamical stability of the PDDE model. Further, we study the entropy evolution and validity of GSL  in sec. 6. Finally, we summarize the conclusions of the work in the last section.
\section{The PDDE model}
In the standard model of cosmology, we have the cosmological constant with the equation of state $\omega_{\Lambda} = -1$. Hence the dark energy density $\rho_{\Lambda}$ and the corresponding pressure satisfies the equation $\rho_{\Lambda} + p_{\Lambda} = 0$. In PDDE model the phantom dark energy density $\rho_D$ and the corresponding pressure $p_D$ satisfies the equation\cite{bouhmadi2015little}
\begin{align}
	\label{eqn:rho1}
	\rho_D +p_D = -\frac{\alpha}{3},
\end{align}
where $\rho_D$ has a specific form given by
\begin{align}
	\label{eqn:rho2}
	\rho_{D}(z) = \rho_{D_0} - \alpha\ln(1+z),
\end{align}
where $z$ is the redshift and $\rho_{D_0}$ is the dark energy density at present. The $\alpha$ is assumed to be a positive constant, and its non-zero value distinguishes this model from the standard $\Lambda$CDM model. Obtaining the equation of state parameter $\omega_D$ from eq. (\ref{eqn:rho1}) and (\ref{eqn:rho2}) is straightforward. 
\begin{align}
	\label{eqn:rho3}
	\omega_D = -\left(1 + \frac{\alpha}{3(\rho_D - \alpha\ln(1+z))}\right).
\end{align}
It is evident from eq. (\ref{eqn:rho3}) that the model mimics quintessence behavior of dark energy ($\omega_D >-1$) for $\alpha < 0$ and reduces to $\Lambda$CDM model when $\alpha = 0$. In the following discussion, we focus on the case, $\alpha > 0$, for which the model shows a phantom behavior ($\omega_{D} < -1$). The non-relativistic matter is assumed to have its kinetic pressure zero, and it satisfies the conservation equation
\begin{align}
	\label{eqn:rho4}
	\dot{\rho_m} + 3H\rho_m = 0,
\end{align}
which on solving gives $\rho_m = \rho_{m_0}(1+z)^{3}$. In this work, we mainly focus on the late-phase cosmology. Hence, we avoid the radiation density as its current value is negligibly small compared to dark matter and dark energy densities. Also, the latest CMBR data obtained by the Planck collaboration suggest that spatial curvature is consistent with a flat universe\cite{aghanim2020planck}. With these observational constraints, the first Friedmann equation takes the form,
\begin{align}
	\label{eqn:H1}
	H^2(z) = H_0^2(\Omega_{m_0}(1+z)^3 + \Omega_{D_0} - \Omega_{pdde}\ln(1+z)).
\end{align}
where $H_0$ is the Hubble parameter at present, $\Omega_{m_0} = 8\pi G\rho_{m_0}/3H_0^2$ is the present value matter density parameter, $\Omega_{D_0} = 8\pi G\rho_{D_0}/3H_0^2$ is the dark energy density parameter and $\Omega_{pdde} = 8\pi G\alpha/3H_0^2$ is the parameter that signifies the PDDE model. From eq. (\ref{eqn:H1}), when $z=0$, $H=H_0$, where $H_0$ is the Hubble parameter at present. In the extreme past, $z\rightarrow\infty$ $(1+z)^3$ dominated over the $\ln(1+z)$ and hence it is conclusive that the non-relativistic matter dominated over the dark energy in the past. In the far future, when $z\rightarrow -1$, matter density tends to zero, and dark energy density tends to infinity. Hence, the Hubble rate diverges in this limit while $\dot{H}$ is a finite constant.
\section{Observational constraints on model parameters}
In PDDE model, there are three independent free parameters $H_0$, $\Omega_{m_0}$ and $\Omega_{pdde}$. The $\Omega_{D_0}$ and $\Omega_{m_0}$ satisfy the equation $\Omega_{m_0} + \Omega_{D_0} = 1.$ Our next aim is to constrain these parameters by confronting the model with the observational data. We adopt Markov Chain Monte Carlo (MCMC) method for the parameter inference\cite{trotta2008bayes}. In this work, we mainly focus on the late phase cosmology and hence we use cosmological data obtained at low redshift. The dataset comprise of Type Ia supernovae data taken from the pantheon sample\cite{scolnic2018complete}, baryon acoustic oscillation data (BAO)\cite{alam2017clustering, de2019baryon, cao2020cosmological, ryan2019baryon} and observational Hubble data (OHD)\cite{Zheng:2016jlq}. The fundamental input for calculating the marginal likelihood of the model parameters is the prior range for each model parameters. We use uniform prior following the ref.\cite{dahmani2023smoothing} for all the parameters that is presented in tab. \ref{tab:prior}. 
The pantheon sample consists of 1048 redshift (z) vs apparent magnitude ($m$) of Type Ia supernovae in the redshift span $0.01\leq z \leq 2.3$\cite{scolnic2018complete}. 
\begin{table}
	\centering
	\caption{Uniform prior chosen for different model parameters}
	\begin{tabular}{ | c | c | }
		\hline
		Parameters & Prior \\ 
		\hline
		$H_0$ & [40, 100]\\  
		$\Omega_{m_0}$ & [0, 1] \\ 
		$\Omega_{pdde}$ & [0, 1] \\
		$M$ & [-20, -18] \\
		\hline
	\end{tabular}
	\label{tab:prior}
\end{table}
\begin{figure}
	\centering
	\includegraphics[width=0.6\linewidth]{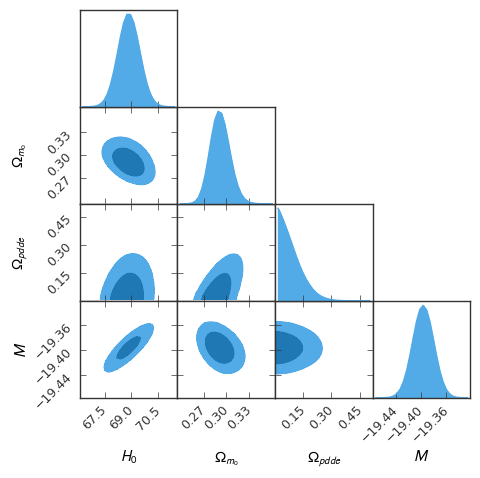}
	\caption{The 2D confidence contour for $68\%$ and $95\%$ probabilities and 1D posterior distribution of the model parameters using SNIa+BAO+OHD dataset.}
	\label{fig:corner}
\end{figure}
To obtain the apparent magnitude from the theory, we use the following expression,
\begin{align}
	\label{eqn:m}
	m(z) = 5\log_{10}\left[\frac{d_L(z)}{Mpc}\right] + M + 25.
\end{align}
Here, $M$ is the absolute magnitude of the Type Ia supernovae and $d_L$ is the luminosity distance that can be related to the expansion rate $H(z)$ as
\begin{align}
	\label{eqn:dl}
	d_L(z) = c(1+z)\int_{0}^{z}\frac{dz'}{H(z')},
\end{align}
where $c$ is the speed of light in vacuum expressed in km/s. The $\chi^2$ for the SNIa data is computed using the expression,
\begin{align}
	\label{eqn:chi1}
	\chi^2_{SNIa} = \sum_{i = 1}^{1048}\frac{\left[m_{th}(z_i,H_0,\Omega_{m_0},\Omega_{pdde}, M) - m_{obs}(z_i)\right]^2}{\sigma(z_i)^2},
\end{align}
where $m_{th}$ and $m_{obs}$ are the apparent magnitude computed using the PDDE model and one that obtained by observation respectively and $\sigma$ is the corresponding error in the measurement.
The BAO measurements include the transverse comoving distance, which is equal the line of sight comoving distance for flat space, is expressed as\cite{Lian:2021tca} 
\begin{align}
	\label{eqn:dm}
	D_M(z) = D_c(z) =\frac{c}{H_0}\int_{0}^{z}\frac{dz'}{h(z')},
\end{align}
and the volume averaged angular diameter distance,
\begin{align}
	\label{eqn:dv}
	D_V(z) = \left[\frac{cz}{H_0}\frac{D_M^2(z)}{h(z)}\right]^{1/3}.
\end{align}
We use BAO data presented in ref.\cite{Lian:2021tca} for the analysis. The $\chi^2$ for the BAO data is obtained using the expression
\begin{align}
	\label{eqn:chi2}
	\chi^2_{BAO} = \sum_{i=1}^{N}\frac{\left[A_{th}(z_i,H_0,\Omega_{m_0},\Omega_{pdde}) - A_{obs}(z_i)\right]^2}{\sigma(z_i)^2},
\end{align}
where $A_{th}$ and $A_{obs}$ are the theoretical and observed value of the physical quantity ($D_M(z_i)$ or $D_V(z_i)$) obtained from the BAO data respectively and $N$ is the number of data points in the BAO measurement. The OHD dataset contain 43 redshift versus Hubble parameter data in the redshift range $0.07\leq z \leq 2.36$\cite{farooq2017hubble, Lian:2021tca}. This data set include the Hubble parameter obtained from cosmic chronometers, radial BAO signals in the distribution of galaxies and BAO signals in the Lyman $\alpha$ forest distribution. The $\chi^2$ for the OHD data can be computed using the expression
\begin{align}
	\label{eqn:chi3}
	\chi^2_{OHD} = \sum_{i=1}^{43}\frac{\left[H_{th}(z_i,H_0,\Omega_{m_0},\Omega_{pdde}) - H_{obs}(z_i)\right]^2}{\sigma(z_i)^2},
\end{align}
where $H_{th}$ and $H_{obs}$ are the Hubble parameter computed theoretically using the PDDE model and the corresponding observed one respectively. Then the total $\chi^2$ is of the form,
\begin{align}
	\label{eqn:chi4}
	\chi^2_{total} = \chi^2_{SNIa} + \chi^2_{BAO} + \chi^2_{OHD}.
\end{align}
The parameter set ($H_0$, $\Omega_{m_0}$, $\Omega_{pdde}$, $M$) that minimize the $\chi^2_{total} = -2\ln \mathcal{L}$, where $\mathcal{L}$ is the likelihood, is considered as the best-fit parameters. The best-fit model parameters are presented in tab.\ref{tab:best}. For the SNIa+BAO+OHD data set, the marginal likelihood of the model parameters are shown in fig. \ref{fig:corner}.
\begin{table}
	\centering
	\caption{The best-fit model parameters of the PDDE model and its uncertainties within $1\sigma$ confidence limit} 
	\begin{tabular}{ | c | c | c | c | c | c | c | }
		\hline
		Data  & $H_0$ & $\Omega_{m_0}$ & $\Omega_{pdde}$ & M \\ 
		\hline
	SNIa+BAO+OHD & $68.86\pm 0.5746$ & $0.291\pm0.011$ & $0.063\pm0.059$ &  $-19.39\pm0.0157$ \\
		\hline
	\end{tabular}
	\label{tab:best}
\end{table}
The minimum $\chi^2$ obtained is $1064.54$ with $\chi^2_{d.o.f} = 0.974$. From tab. \ref{tab:best}, it is clear that the estimated value of $H_0$ and $\Omega_{m_0}$ and $M$ are consistent with the standard values obtained in the literature\cite{aghanim2020planck, riess1998observational}. The phantom dark energy density $\Omega_{pdde}$ that characterize the PDDE model is one order of magnitude less than the matter density as expected. we don't expect much deviation from the $\Lambda$CDM predictions. Also the large value of the standard deviation which is comparable to the best-fit value indicate that the significance of non-zero value of $\Omega_{pdde}$ is $\sim 1\sigma$, which is statistically less significant.

\section{Evolution of cosmographic parameters}
\begin{figure}
	\centering
	\includegraphics[width=0.6\linewidth]{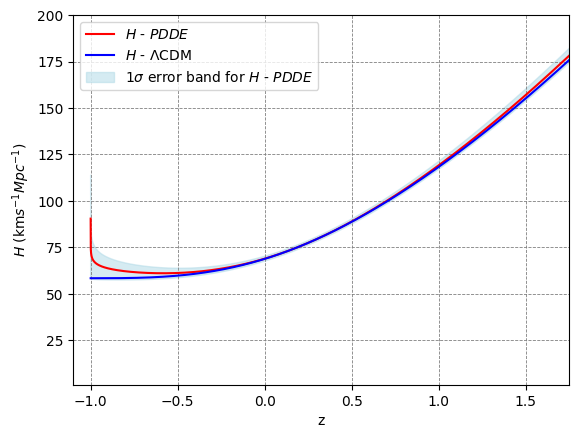}
	\caption{Evolution of Hubble parameter ($H$) with redshift ($z$) is plotted for PDDE model with error band and $\Lambda$CDM using the best-fit model parameters obtained for the SNIa+BAO+OHD data combination.}
	\label{fig:H1}
\end{figure}
In this section, we examine the evolution of various cosmographic parameters to understand how well the PDDE model explain the background evolution of the universe. The parameter of fundamental importance in cosmology is the Hubble parameter. The equation for Hubble parameter is presented in eq. (\ref{eqn:H1}) and its evolution against redshift is depicted in fig. \ref{fig:H1}.
The evolution of Hubble parameter in PDDE model is almost similar to the $\Lambda$CDM model in the past and at present. However, its evolution shows a considerable difference in the asymptotic future. The $\Lambda$CDM model provides a de Sitter  solution in the far future evolution while there exists a big-rip singularity where Hubble parameter blows up for the PDDE model. However this event is going to happen in the infinite future.

\begin{figure}
	\centering
	\includegraphics[width=0.6\linewidth]{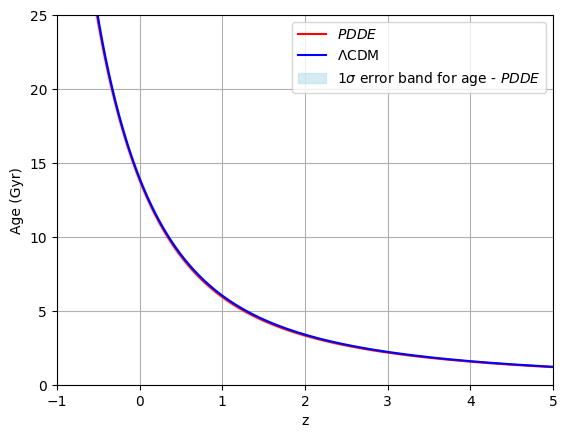}
	\caption{Age of the universe at each redshift ($z$) is plotted for PDDE model with error band and $\Lambda$CDM using the best-fit model parameters obtained for the SNIa+BAO+OHD data combination.}
	\label{fig:age}
\end{figure}

The age of the universe as a function of scale factor can be obtained as follows. The Hubble parameter, by definition, is $H(a) = \dot{a}/a$, on rearranging, we get $dt/da = (aH(a))^{-1}$. Hence, the age of the universe at any scale factor $a$ is expressed as 
\begin{align}
	\label{eqn:age}
t_a - t_B = \int_{0}^{a} \frac{1}{aH(a)}da.
\end{align}
Here, $t_a$ is the age of the universe at scale factor $a$ and $t_B$ is the age of the universe at Big Bang which is assumed to be zero. The universe's age with respect to redshift is plotted in fig. \ref{fig:age}.
The present age of the universe is the age at $z=0$, the estimated age is $13.86\pm 0.27$ Gyr. The age of the universe computed for the $\Lambda$CDM model is $13.78$ Gyr and the one that is obtained from the CMB data assuming the standard $\Lambda$CDM model\cite{aghanim2020planck} is $13.79$ which are within the $1\sigma$ error band of  the age predicted by the PDDE model. 

\begin{figure}
	\centering
	\includegraphics[width=0.6\linewidth]{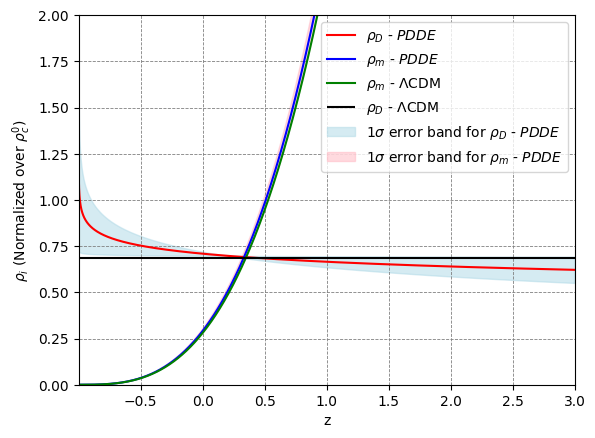}
	\caption{Evolution of matter density and dark energy density against redshift are plotted for PDDE model with error band and $\Lambda$CDM using the best-fit model parameters obtained for the SNIa+BAO+OHD data combination.}
	\label{fig:density}
\end{figure}

Following the individual conservation equation satisfied by matter and dark energy, the evolution of matter energy density is similar to that of the $\Lambda$CDM model. The evolution of phantom dynamical dark energy is different from the $\Lambda$CDM model. In $\Lambda$CDM model the dark energy is just the cosmological constant while phantom dark energy is dynamical. However, the phantom dynamical dark energy density become indistinguishable from the cosmological constant at present. The evolution of matter energy density and dark energy density for PDDE model and $\Lambda$CDM model are presented in fig. \ref{fig:density}. The matter density and dark energy density in the $\Lambda$CDM model are within the 1$\sigma$ error band of the PDDE model throughout the late phase evolution of the universe.

The deceleration parameter ($q$) characterizes the accelerating or decelerating expansion of the FRW universe. It is expressed as
\begin{equation}
	\label{eqn:q}
	q = -1-\frac{\dot{H}}{H^2}.
\end{equation}
It is convenient to express the eq. \ref{eqn:H1} in terms of dimensionless Hubble parameter $h = H/H_0$ as
\begin{equation}
	\label{eqn:q0}
	q = -1-\frac{1}{2h^2}\frac{dh^2}{dx},
\end{equation}
where over dot represents the derivative with respect to cosmic time. Substituting for $h^2$ from eq. (\ref{eqn:H1}), we obtain the deceleration parameter that varies with scale factor as
\begin{equation}
	\label{eqn:decc}
	q=\frac{\Omega_{m_0}a^{-3}-2\Omega_{D_0}-\Omega_{pdde}(2\ln{a}+1)}{2\Omega_{m_0}a^{-3}+2\Omega_{D_0}+2\Omega_{pdde}\ln{a}}.
\end{equation}

\begin{figure}
	\centering
	\includegraphics[width=0.6\linewidth]{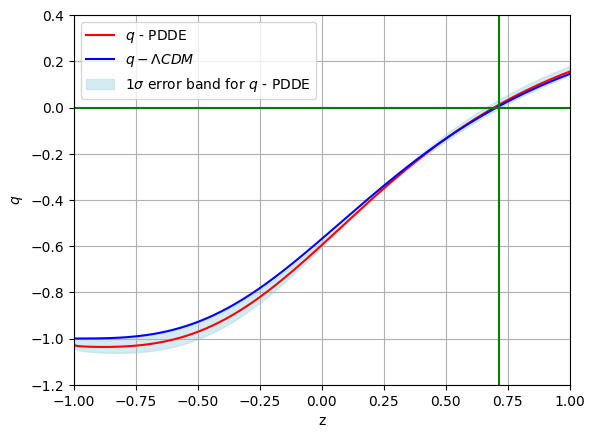}
	\caption{Progress  of deceleration parameter ($q$) of the universe with respect to redshift ($z$) is plotted for PDDE model with error band and $\Lambda$CDM using the best-fit model parameters obtained for the SNIa+BAO+OHD data combination. The intersection of the green line represent the present value of the deceleration parameter ($q_0$).}
	\label{fig:q}
\end{figure}

The progress of deceleration parameter with respect to redshift is plotted in fig. \ref{fig:q}. The positive value of $q$ indicating a decelerating universe and the universe is accelerating if the value of $q$ is negative. From the fig. \ref{fig:q}, it is evident that the universe made a decelerating to accelerating transition. The transition redshift is computed as $z_T = 0.69\pm 0.03$ while the transition redshift computed for the $\Lambda$CDM model using the same data set combination, SNIa+BAO+OHD is $z_T = 0.70$. It is clear that the value obtained for $\Lambda$CDM model is within the $1\sigma$ error band of the value obtained using the PDDE model. The present value of the deceleration parameter is $q_0 = -0.59\pm 0.01$, showing the the universe is accelerating at present. The PDDE model is successful in explaining the recent deceleration to acceleration transition that occurred in the recent past and also the observed present acceleration of the universe.

\begin{figure}
	\centering
	\includegraphics[width=0.6\linewidth]{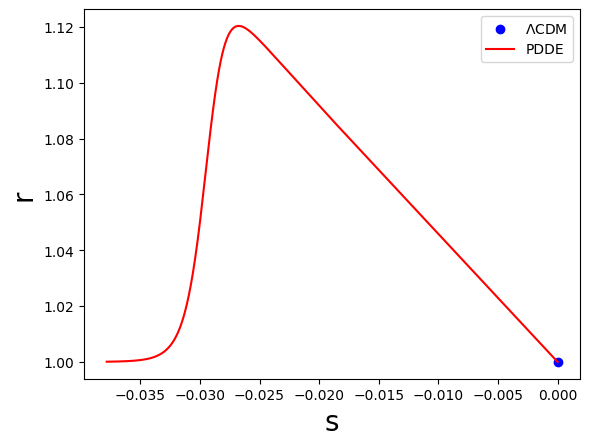}
	\caption{Statefinder trajectory is plotted for PDDE model with the model parameters obtained using the SNIa+BAO+OHD data combination. The blue dot represents the $\Lambda$CDM fixed point.}
	\label{fig:rs}
\end{figure}

We compute the statefinder diagnostic pair, the jerk parameter ($r$) and snap parameter ($s$)\cite{sahni2003statefinder} to characterize the dark energy models and distinguish between them. The statefinder parameters, $r = 1$ and $s =0$ is a fixed point for the standard $\Lambda$CDM model. The jerk parameter is defined as
\begin{align}
	\label{eqn:r}
	r = \frac{1}{aH^3}\frac{d^3a}{dt^3}.
	\end{align}
It is convenient to express the $r$ parameter in terms of reduced Hubble parameter $h = H/H_0$, we obtain
\begin{align}
	\label{eqn:r1}
	r = \frac{1}{2h^2}\frac{d^2h^2}{dx^2} + \frac{3}{2h^2}\frac{dh^2}{dx} + 1.
\end{align}
The snap parameter is defined as
\begin{align}
	\label{eqn:s0}
	s = \frac{1}{aH^4}\frac{d^4a}{dt^4}.
\end{align}
The $s$ parameter can also be written in terms of $r$ parameter and $q$ parameter as
\begin{align}
	\label{eqn:s}
	s = \frac{r-1}{3(q-\frac{1}{2})}.
\end{align}
On expressing $s$ in terms of reduced Hubble parameter, we get
\begin{align}
	\label{eqn:s1}
	s = -\frac{\frac{1}{2h^2}\frac{d^2h^2}{dx^2} + \frac{3}{2h^2}\frac{dh^2}{dx}}{\frac{3}{2h^2}\frac{dh^2}{dx}+\frac{9}{2}},
\end{align}
where we have changed the variable from $a$ to $x=\ln a$. On substituting the expression of $h^2$ from eq. (\ref{eqn:H1}) in eq. (\ref{eqn:r1}) and (\ref{eqn:s1}), we obtained the evolution of $r$ and $s$ parameter in PDDE model as
\begin{equation}
	\label{eqn:rp}
	r=\frac{3\Omega_{pdde}}{2\Omega_{m_0}a^{-3}+2\Omega_{D_0}+2\Omega_{pdde}\ln{a}}+1,
\end{equation}
and
\begin{equation}
	\label{eqn:sp}
	s= -\frac{\Omega_{pdde}}{\Omega_{pdde}+3\Omega_{D_0}+3\Omega_{pdde}\ln{a}}.
\end{equation}
From eq. (\ref{eqn:rp}) and (\ref{eqn:sp}), it is clear that the $r$ and $s$ parameters are dynamical quantities. Also, the parameters $(r, s) = (1, 0)$ is a fixed point when $\Omega_{pdde} = 0$. The r-s trajectory is shown in fig. \ref{fig:rs}. It shows that $r>1$ and $s<0$ until the trajectory reaches the $\Lambda$CDM fixed point, confirms the phantom nature of dark energy density. The present values of the $r$ and $s$ are computed as $(r, s) = (1.09, -0.22)$ showing that the PDDE model is distinguishable from the $\Lambda$CDM model at present. 
\section{Dynamical system analysis}
Dynamical system analysis is a useful tool in cosmology to extract asymptotic stability of a cosmological model\cite{wainwright1997dynamical, coley2003dynamical}. Fundamental input in dynamical system analysis is the choice of appropriate dynamical variables.  In this study, we consider two dimensionless dynamical variables $u$ and $v$, defined as
\begin{align}
	\label{eqn:u}
	u = \frac{\rho_m}{3H^2} \hspace{1.5cm} v = \frac{p_D}{3H^2}.
\end{align}
System of autonomous differential equations can be formulated by taking derivative of $u$ and $v$ with respect to the cosmic variable $x =\ln a$, we obtain 
\begin{align}
	\label{eqn:du}
	\frac{du}{dx} = 3uv = f(u,v), \\
	\label{eqn:du1}
	\frac{dv}{dx} = 3-3u+6v+3v^2 = g(u,v).
\end{align}
The critical points $(u_c, v_c)$ are obtained  by solving the equations $du/dx = 0$ and $dv/dx = 0$, we get
\begin{equation}
\begin{aligned}
	(u_c, v_c) &= (1, 0), \\
	(u_c, v_c) &= (0, -1).
\end{aligned}
\end{equation}
The stability of the dynamical system near the critical points can be understood by linearizing the system by considering small variation around the critical points,
\begin{align}
	\label{eqn:u1}
	u\rightarrow {u_c} + \delta u \hspace{1.5cm} v \rightarrow {v_c} + \delta v,
\end{align}
that satisfy the matrix equation
\begin{gather}
	\label{eqn:matrix}
	\begin{pmatrix} \delta u'  \\ \delta v' \end{pmatrix}
	=
	\begin{pmatrix}
		\left(\frac{\partial f}{\partial u}\right)_{u_c, v_c} &
		\left(\frac{\partial f}{\partial v}\right)_{u_c, v_c} \\
		\left(\frac{\partial g}{\partial u}\right)_{u_c, v_c} &
		\left(\frac{\partial g}{\partial v}\right)_{u_c, v_c} 
	\end{pmatrix}
\begin{pmatrix} \delta u  \\ \delta v \end{pmatrix}
\end{gather}
The Jacobian matrix given in eq.\ref{eqn:matrix} evaluated at the critical points (1, 0) and (0, -1) respectively are
\begin{gather}
J_{(1, 0)} = \begin{pmatrix}
	0 &
	3 \\
	-3 &
	6
\end{pmatrix}
\hspace{1cm}
J_{(0, -1)} = \begin{pmatrix}
	-3 &
	0 \\
	-3 &
	0
\end{pmatrix}
\end{gather}
\begin{figure}
	\centering
	\includegraphics[width=0.6\linewidth]{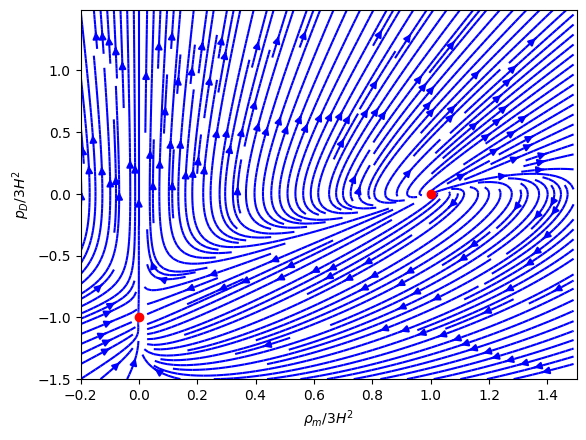}
	\caption{Phase space trajectory in the ($\rho_m/3H^2$) - ($p_D/3H^2$) plane for the PDDE model. The red dots represent the critical points.}
	\label{fig:phase}
\end{figure}
The stability of the critical points are determined by the eigenvalues of the matrix J. If both of the eigenvalues are negative, the critical point is a future attractor which is asymptotically stable. All the trajectories near the critical point approach that point. If both the eigenvalues are positive, the critical point is a past time attractor or an unstable equilibrium point. All the trajectories near the critical point will be repelled from the point. If one eigenvalue is positive and other negative, the critical point is a saddle point. Some trajectories will be attracted to the critical point while others will repel. If one of the eigenvalue is positive and other zero, the critical point is unstable. The positive eigenvalue guarantee that there exist at least one unstable direction. If one of the eigenvalue is zero and other is negative, the linear stability theory can not explain the stability of the critical point. If both eigenvalues are complex numbers of the form $\beta+i\gamma$ and $\beta-i\gamma$ with $\beta > 0$ and $\gamma \neq 0$, the critical point is an unstable spiral. If $\beta < 0$ and $\gamma \neq 0$, the critical point is a stable spiral. If $\beta = 0$, the critical point is called a center and the solutions are oscillatory\cite{bohmer2017dynamical}. Diagonalizing $J_{(1, 0)}$ and $J_{(0,-1)}$, we obtained the eigenvalues corresponds to the critical points $(1,0)$ and ($0, -1$) as $(3, 3)$ and $(-3, 0)$ respectively. The critical point ($\rho_m/3H^2$, $p_D/3H^2$) = ($1,0$) corresponds to a fixed point in the matter dominated phase. Since both the eigenvalues are positive, it represents an unstable equilibrium point.  The critical point ($\rho_m/3H^2$, $p_D/3H^2$) = ($0,-1$) corresponds to a fixed point in the far future where matter density is zero and the universe is completely occupied with vacuum energy. The stability of this critical point cannot be determined by analysing the eigenvalue as the point is non-hyperbolic. The phase space portrait of the dynamical system is shown in fig. \ref{fig:phase}.
From fig. \ref{fig:phase}, it is evident that the critical point $(1, 0)$ is an unstable equilibrium point so that all the trajectories in the neighbourhood of this point are repelled out from the point. However, the stability of the critical point $(0, -1)$ can not be inferred from the phase space plot. Here, the linear stability analysis fails to explain the stability of the critical point. Hence, we use the centre manifold theory to simplify the system by reducing the dimensionality\cite{wiggins2003introduction, bohmer2017dynamical, bahamonde2018dynamical}. 

The model has a critical point at $(0,-1)$ with eigen values $(-3, 0)$. To use the central manifold theory, we shift the point to the origin or equivalently transform the eq. (\ref{eqn:du}) and (\ref{eqn:du1}) into a set of new variables given by
\begin{equation}
	\label{eqn:N}
	U = u, \hspace{1.5cm}  V = v+1.
\end{equation}
Then, the the eq. (\ref{eqn:du}) and (\ref{eqn:du1}) can be re-written as
\begin{align}
	\label{eqn:N1}
	\frac{dU}{dx} = 3UV-3U,\\
	\label{eqn:N2}
	\frac{dV}{dx} = 3V^2 - 3U.
\end{align}
The Jacobian matrix for the system can be expressed as
\begin{gather}
		J = \begin{pmatrix}
		3V-3 &
		3U \\
		-3 &
		6V
	\end{pmatrix}.
\end{gather}
The Jacobian matrix for the critical point (0,0) is 
\begin{gather}
	J|_{U =0, V=0} = \begin{pmatrix}
		-3 &
		0 \\
		-3 &
		0
	\end{pmatrix}.
\end{gather}
Now, we construct another set of variables $(X, Y)$ defined as
\begin{gather}
	\begin{pmatrix}
		X 
		\\
		Y
	\end{pmatrix}
= S^{-1}
	\begin{pmatrix}
	U 
	\\
	V
\end{pmatrix},
\end{gather}
where $S$ is the inverse of the diagonalizing matrix constructed out of the eigenvectors of the matrix $J|_{U=0, V=0}$. The matrix S obtained is
\begin{gather}
	S = \begin{pmatrix}
		0 &
		1 \\
		1 &
		1
	\end{pmatrix}.
\end{gather}
Now, it is possible to express $U$ and $V$ in terms of new dynamical variables $X$ and $Y$ as
\begin{align}
	\label{eqn:N3}
	V = X + Y, \hspace{1.5cm} U = Y.
\end{align}
Then, the eq. (\ref{eqn:N1}) and (\ref{eqn:N2}) can be expressed in terms of the new variables $X$ and $Y$ as
\begin{align}
	\label{eqn:N4}
	\dot{X} = 3X^2 + 3XY\\
	\label{eqn:N5}
	\dot{Y} = 3XY+3Y^2-3Y,
\end{align}
which according to central manifold theory can be represented as 
\begin{align}
	\label{eqn:N6}
	\dot{X} = AX + F(X,Y)\\
	\label{eqn:N7}
	\dot{Y} = BY + G(X,Y),
\end{align}
Comparing eq. (\ref{eqn:N4}) and (\ref{eqn:N6}), we obtain $A =0$ and $F(X,Y) = 3X^2 + 3XY$. Similarly, from eq. (\ref{eqn:N5}) and (\ref{eqn:N7}) we obtain $B = -3$ and $G(X,Y) = 3XY+3Y^2$. Assuming $Y = h(X)$, for $h(X)$ to be the centre manifold, it satisfies the quasilinear partial differential equation
\begin{align}
	\label{eqn:diff}
	\frac{dh(X)}{dX}\left[AX+F(X,h(X))\right] - Bh(X) - G(X,h(X)) = 0
\end{align}
In order to proceed with determining the centre manifold, it is customary to suppose the expansion for $h(X)$ of the form
\begin{align}
	\label{eqn:hx}
	h(X) = a_2X^2 + a_3X^3 + \mathcal{O}(X^4).
\end{align} 
Substituting eq. (\ref{eqn:hx}) into eq. (\ref{eqn:diff}), we obtain
\begin{align}
	\label{eqn:hx1}
	(2a_2X + 3a_3X^2)(3X^2+3X(a_2X^2 + a_3X^3)) - 3X(a_2X^2 + a_3X^3) \nonumber\\ - 3(a_2X^2 + a_3X^3)^2 + 3(a_2X^2 + a_3X^3) = 0.
\end{align} 
For the eq. (\ref{eqn:hx1}) to holds, the coefficient of each powers of X must be zero. Then, we get the coefficients $a_2 = 0$, $a_3 = 0$ and hence $h(X) = \mathcal{O}(X^4).$ Then, the dynamical system restricted to centre manifold is expressed as
\begin{align}
	\label{eqn:dy}
	\dot{X} = 3X^2 + \mathbf{O}(X^5).
\end{align}
From eq. (\ref{eqn:dy}), it can be concluded that for sufficiently small $X$, $X \rightarrow 0$, the system is unstable. Hence the critical point $(0,0)$ is unstable.
  The analysis shows that there is no stable equilibrium point exist in any evolutionary stage of the universe. Hence, we conclude that the universe is dynamically unstable within the PDDE model. It  should be noted that the universe is dynamically stable and attain a de Sitter evolution in the far future within the $\Lambda$CDM model. Hence, we conclude that, even though the present model alleviate the Hubble tension to some extent, the model is not dynamically stable and hence may not be considered as a potential alternate to $\Lambda$CDM model.
\section{Evolution of horizon entropy and GSL}
\begin{figure}
	\centering
	\includegraphics[width=0.6\linewidth]{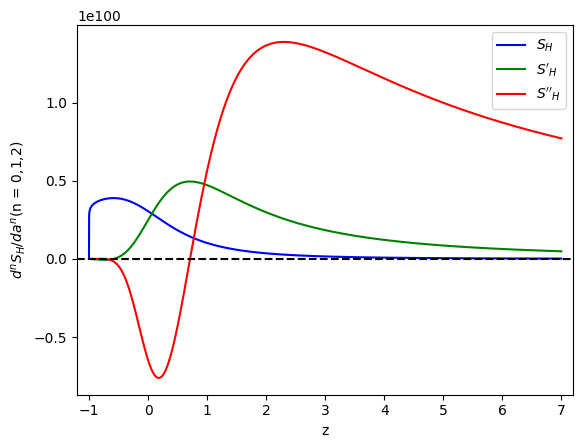}
	\caption{Evolution of horizon entropy ($S_H$), first and second derivative ($S'_H$ and $S''_H$) of horizon entropy against redshift.}
	\label{fig:entropy}
\end{figure}

It was Bekenstein and Hawking who showed that the entropy of black holes are directly proportional to  area of their event Horizon\cite{bekenstein1973black, bekenstein1974generalized, hawking1976black}. The event horizon entropy ($S_{EH}$) is given by 
\begin{align}
	\label{eqn:EH}
	S_{EH} = \frac{A_{EH}}{4l_p^2}k_B. 
\end{align}
where $A_{EH}$ is the area of the event horizon, $l_p$ is the Planck length and $k_B$ is the Boltzmann constant. Later, Gibbons and Hawking showed that cosmological horizon also possess entropy proportional to area\cite{gibbons1977cosmological, jacobson1995thermodynamics}. Instead of event horizon, Hubble horizon is mostly used for the estimation of horizon entropy because it provides the observable boundary at the present time. Hence, the area-entropy relation for the observable universe can be expressed as
\begin{align}
	\label{eqn:SH}
	S_H = \frac{A_{H}}{4l_p^2}k_B,
\end{align}
where $A_{H} = 4\pi c^2/H^2$ is the area of the Hubble horizon. According to Generalized second law of thermodynamics (GSL), entropy of the universe (horizon entropy + entropy of its interior) must be a never decreasing function of time\cite{PhysRevLett.84.2072, bekenstein1974generalized},
\begin{align}
	\label{eqn:GSL}
	\dot{S} = \dot{S}_H + \dot{S}_m \geq 0,
\end{align}
where $S_m$ is the entropy contribution from anything present inside the Hubble horizon and the overdot denote the derivative with respect to the cosmic time. In general, the $S_m$ include entropy contribution from baryon matter ($S_b \sim 10^{81}k_B$), dark matter ($S_{dm} \sim 10^{88\pm 1}k_B$), photons ($S_{rad} \sim 10^{89}k_B$), relic neutrinos ($S_{\nu} \sim 10^{89}k_B$), relic gravitons ($S_{grav} \sim 10^{87}k_B$), stellar black holes ($S_{SBH} \sim 10^{97}k_B$) and supermassive black holes ($S_{SMBH} \sim 10^{104}k_B$). However, the horizon entropy ($S_H \sim 10^{122}k_B$) is very much higher than that of the entropy contribution due to any of the cosmic components mentioned above\cite{egan2010larger}. Hence, we consider only the horizon entropy for the following discussions. The horizon entropy is obtained by substituting the expression of horizon area into the eq. (\ref{eqn:SH}), we obtain
\begin{align}
	\label{eqn:SH0}
	S_H = \frac{\pi c^2}{l_p^2 H^2}k_B.
\end{align}
where the evolution of the Hubble parameter is given by eq. (\ref{eqn:H1}). The derivative of horizon entropy with respect to redshift is obtained as
\begin{align}
	\label{eqn:SH1}
	S'_H = \frac{\pi c^2 k_B}{l_p^2H_0^2}\left(\frac{3\Omega_{m_0}a^{-4} - \Omega_{pdde}a^{-1}}{h^4}\right).
\end{align}
To check the convexity condition of entropy, we also obtained the expression for second derivative of entropy with respect to scale factor, 
\begin{align}
	\label{eqn:SH2}
	\!\!\!\!\! S''_H = \frac{\pi c^2 k_B}{l_p^2H_0^2}\left[-\frac{1}{h^4}\left(12\Omega_{m_0}a^{-5} - \Omega_{pdde} a^{-2}\right) + \frac{2}{h^6}\left(3\Omega_{m_0}a^{-4} - \Omega_{pdde}a^{-1}\right)^2\right].
\end{align}
The evolution of $S_H$ (blue curve), $S'_H$ (green curve) and $S''_H$ (red curve) are depicted in fig. \ref{fig:entropy}.
\begin{figure}
	\centering
	\begin{subfigure}{0.4\textwidth}
		\centering
		\includegraphics[width=\textwidth]{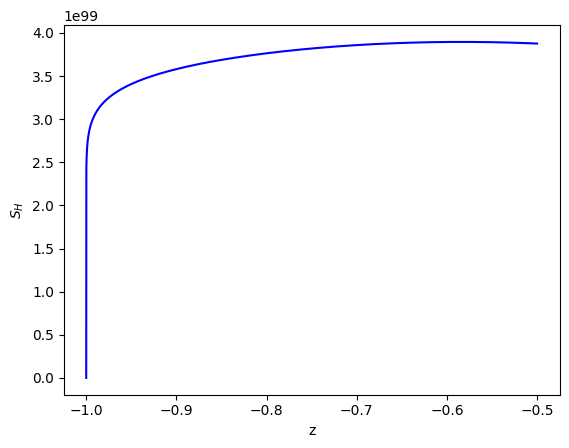}
		\label{fig:y equals x}
	\end{subfigure}
	\hfill
	\begin{subfigure}{0.4\textwidth}
		\centering
		\includegraphics[width=\textwidth]{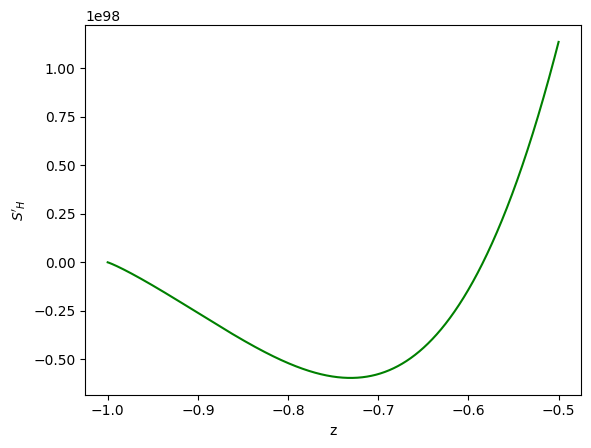}
		\label{fig:three sin x}
	\end{subfigure}
	\\
	\begin{subfigure}{0.4\textwidth}
		\centering
		\includegraphics[width=\textwidth]{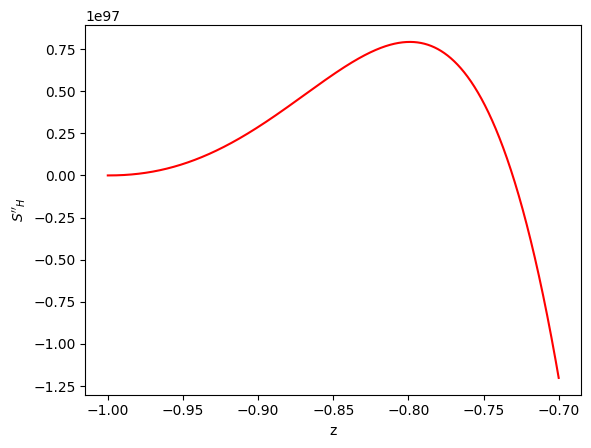}
		\label{fig:five over x}
	\end{subfigure}
	\caption{The $S_H$ (blue curve), $S'_H$  (green curve) and $S''_H$ (red curve) are plotted against the redshift and zoomed to explicitly show the violation of generalized second law of thermodynamics}
	\label{fig:zoom}
\end{figure}
From fig. \ref{fig:entropy}, it is conclusive that entropy decreases with respect to scale factor in the far future evolution. However, it is not evident from the $S'_H$ and $S''_H$ curves. To show this behaviour more explicitly, we zoomed the individual curve in the region between $z= -1$ and $z = -0.5$, the plots are shown in fig. \ref{fig:zoom}.
 From fig. \ref{fig:zoom}, it is clear that $dS_H/da$ has both positive and negative values. But, according to GSL, $dS_H/da \geq 0$ throughout the evolution. Negative value of $dS_H/da$ in the later stage of evolution is a clear violation of GSL. Furthermore, it is well known that any isolated system evolve towards a state of maximum entropy. In other words, the entropy should be a convex function of scale factor, that is
\begin{align}
	\label{eqn:SH3}
	S''_H < 0,
\end{align}
at least in the later stage of evolution. Fig. \ref{fig:zoom} shows that the convexity condition is not satisfied for a range of redshift in the late accelerating phase. In this region the second $S''_H$ attain a maximum positive value and then approaching to zero from above. The entropy is not bounded from below. This implies that entropy evolution of the universe within the PDDE model doesn't behaves like an ordinary macroscopic system. In conclusion, the PDDE model violate the generalized second law of thermodynamics and it doesn't consistent with the entropy maximization.
\section{Conclusion}
In the present work, we examine the dynamical stability of the PDDE model and its thermodynamic consistency. The PDDE model, also termed as little sibling of the big rip is a new event which is smoother than the existing big rip singularities within the phantom dark energy models. This event happen at $a\rightarrow \infty$, when the scalar curvature explodes. When this event is reached, the scale factor and the Hubble parameter diverges while the derivative of Hubble parameter remains finite. The model behaves like $\Lambda$CDM at present. However, the universe wouldn't be asymptotically de Sitter as in the $\Lambda$CDM model\cite{bouhmadi2015little}. 
We performed parameter inference of the model based on MCMC using a string of observations SNIa+BAO+OHD. Our analysis shows that the model is in good agreement with the observational data. The best-fit values of the model parameters obtained are $H_0 = 68.86\pm 0.5746$, $\Omega_{m_0} = 0.291\pm0.011$ and $\Omega_{pdde} = 0.063\pm0.059$. These values are consistent with the values reported by Bouali et al.\cite{bouali2019cosmological}. It should be noted that the standard deviation obtained for the parameter $\Omega_{pdde}$ is almost equal to the best-fit value. This implies that the data prefer the non-zero value of $\Omega_{pdde}$ at a significance of $\sim 1\sigma$ which is statistically less significant.

The age of the universe is computed as $13.86\pm 0.27$ GYr which is consistent with the standard model prediction. Evolution of matter density in the PDDE model is same as that in the $\Lambda$CDM model while the dark energy density is not just a cosmological constant instead varying with time. The present value of the deceleration parameter is negative showing that universe is currently accelerating. The decelerating universe made transition into accelerating phase in the recent past at a redshift $z_T = 0.69\pm 0.03$. The statefinder parameter distinguished the PDDE model from the $\Lambda$CDM model and r-s trajectory reveals the phantom nature of dark energy density.

We performed the dynamical system analysis to test the dynamical stability of the PDDE model. On solving autonomous coupled differential equation satisfied by the dynamical variables, $\rho_m/3H^2$ and $p_D/3H^2$, we obtained two critical points ($1, 0$) and ($0, -1$). The point ($1, 0$) is a fixed point in the matter dominated phase and ($0, -1$) is a fixed point in the late accelerating phase. The eigenvalues obtained by diagonalizing the Jacobian matrices are ($3, 3$) and ($-3, 0$) for the critical points ($1, 0$) and ($0, -1$) respectively. The fixed point ($1, 0$) is an unstable equilibrium point and ($0, -1$) is saddle point which is also unstable. In conclusion, the universe is dynamically unstable in the matter dominated epoch and late accelerating epoch within the PDDE model.

Analysis based on entropy evolution shows that the present model violates the generalized second law of thermodynamics in the far future evolution. Further the entropy is not bounded from below for a range of redshift in the future evolution and hence the model doesn't satisfy the entropy maximization. Universe doesn't behaves like an ordinary macroscopic system within the PDDE model.

In summary, despite the present model is in good agreement with the observational data and a potential candidate to smoothen the Hubble tension, the model is not dynamically stable and thermodynamically consistent. Our analysis shows that the PDDE model may not be considered as a potential alternate to standard model of cosmology, the $\Lambda$CDM. In this work, we considered the dark matter and dark energy which are not interacting each other. In more general scenario, the dark matter and dark energy can interact each other. In this regard one can check the dynamical stability of the interacting phantom dynamical dark energy model.

\end{document}